\documentclass[12pt]{iopart}

\usepackage{amssymb}
\usepackage{latexsym}
\usepackage{euscript}
\usepackage{graphics}
\usepackage[all]{xy}

\usepackage{graphicx}% Include figure files
\usepackage{dcolumn}% Align table columns on decimal point
\usepackage{bm}% bold math

\newcommand{\be}{\begin{equation}}
\newcommand{\ee}{\end{equation}}
\newcommand{\ba}{\begin{eqnarray}}
\newcommand{\ea}{\end{eqnarray}}
\newcommand{\baa}{\begin{eqnarray*}}
\newcommand{\eaa}{\end{eqnarray*}}
\newcommand{\bb}{\begin{thebibliography}}
\newcommand{\eb}{\end{thebibliography}}

\renewcommand{\bi}[1]{\bibitem{#1}}
\newcommand{\lab}[1]{\label{#1}}
\newcommand{\re}[1]{(\ref{#1})}

\renewcommand\t{\tilde}

\eqnobysec

\begin{document}
\title[Para-Krawtchouk polynomials]{Para-Krawtchouk polynomials on a bi-lattice and a quantum spin chain with perfect state transfer}
\author{Luc Vinet $^1$ and Alexei Zhedanov$^2$}
\address{$^1$ Centre de recherches math\'ematiques
Universit\'e de Montr\'eal, P.O. Box 6128, Centre-ville Station,
Montr\'eal (Qu\'ebec), H3C 3J7}
\address{$^2$ Institute for Physics and Technology,
R.Luxemburg str. 72, 83114 Donetsk, Ukraine}

\begin{abstract}
Analogs of Krawtchouk polynomials defined on a bi-lattice are
introduced.  They are shown to provide a (novel) spin chain with
perfect transfer. Their characterization is given as well as their
connection to the quadratic Hahn algebra
\end{abstract}


%\keywords{}

%\end{titlepage}

%\maketitle

\vspace{15mm}

\section{Introduction}
\setcounter{equation}{0} Progress in exactly solving physical
models is often tied to advances in the theory of orthogonal
polynomials and special functions (OPSF). We here present one such
instance. In looking for spin chains that admit perfect state
transfer (PST), we have found a new remarkable family of
orthogonal polynomials defined on a bi-lattice (i.e. a lattice formed from the union of two regular lattices). We here report on these findings.

The transfer of quantum state from one location to another is of clear importance in quantum information. It is at the root of exchanges between individuals or between components of an eventual quantum computer. It is also fundamental in the development of quantum algorithms. Spin chains, that is one-dimensional coupled systems of spins, are being used in the design of quantum wires that will effect this transport \cite{Albanese}. These models have the merit of avoiding the need of external control as the dynamics of the chain is responsible for the transmission. An issue of interest is the efficiency of these wires. Ideally, one would wish for the transfer to be perfect, in other words, that the probability be one of finding the input state as output at some time. It has been shown  that this can be achieved in various cases by properly engineering the couplings between the spins \cite{Albanese}, \cite{Kay}.

The simplest systems in which this can be realized are $XX$ spin chains with nearest neighbor interactions and inhomogeneous couplings. Many aspects of PST are revealed by focusing on the one-excitation dynamics. Under this restriction, the $XX$-Hamiltonian takes the form of a Jacobi or tridiagonal Hermitian matrix $J$. Such matrices are intimately connected to orthogonal polynomials as they entail 3-term recurrence relations. Moreover, the conditions for PST to occur can be expressed as conditions on their eigenvalues. As a matter of the fact, PST requires that the differences between neighboring eigenvalues of $J$ be odd integers (up to a scale factor). The matrix $J$ contains the complete information on the couplings of the full Hamiltonian. Determining $XX$ spin chain chains with PST, therefore amounts to an inverse spectral problem that is to the problem of constructing Jacobi matrices $J$ that have given sets of eigenvalues satisfying the PST conditions. Algorithms have been developed to that end by exploiting the underlying connection with the theory of orthogonal polynomials (OPs).

It is of practical interest to identify spin chains with PST that admit exact solutions in terms of special functions. This obviously allows an analytical study of the dynamics. The interest is doubled if the examination of spin chains with PST leads to the full characterization of orthogonal polynomials that had eluded attention. The present paper is in that vein.  We purported to determine the $XX$ spin chains with PST corresponding to one-excitation spectra of bi-lattice form. In so doing, we found a new family of OPs that we have called para-Krawtchouk polynomials. We have studied their properties and found that they share many of the features of classical OPs. 

The rest of the paper is organized as follows.   In Section 2, we review the relation between $XX$ spin chains with PST and orthogonal polynomials. In Section 3, we present a new chain of that type which is amenable to an analytic  treatment. It corresponds to a one-excitation energy spectrum of bi-lattice form and is valid for a chain where the number $N+1$ of sites is even. The coupling coefficients are the recurrence parameters for the para-Krawtchouk polynomials. These turn out to form a one-parameter generalization of the Krawtchouk polynomials which are related to the PST-model first presented in \cite{Albanese}. Section 4 and 5 are devoted to the characterization of the para-Krawtchouk polynomials. In Section 4 an explicit expression is given in terms of the complementary Bannai-Ito polynomials which arose in recent studies of Dunkl or -1 polynomials. In Section 5, their relations with the quadratic Hahn algebra is established. In Section 6, equipped with the appropriate mathematical tools developed in Sections 4 and 5, we appeal to the Christoffel transform to obtain in the case of an odd number $N+1$ of states, the chain with PST which is again associated to a bi-lattice  spectrum.

\section{Orthogonal polynomials and quantum spin chains with perfect state transfer}
\setcounter{equation}{0} Consider the $XX$ spin chain with
Hamiltonian \be H=\frac{1}{2} \: \sum_{l=0}^{N-1}
J_{l+1}(\sigma_l^x \sigma_{l+1}^x +  \sigma_l^y \sigma_{l+1}^y) +
\frac{1}{2} \: \sum_{l=0}^N B_l(\sigma_l^z +1), \lab{H_def} \ee
where $J_l>0$ are the constants coupling the sites $l-1$ and $l$
and $B_l$ are the strengths of the magnetic field at the sites $l$
($l=0,1,\dots,N$). This operator acts on $\mathbb{C}^{2 \otimes (N+1)}$ The symbols $\sigma_l^x, \: \sigma_l^y,\:
\sigma_l^z$ stand for the Pauli matrices, which act as follows on the standard basis 
$\{| 0 \rangle, |1 \rangle\}$ of $\mathbb{C}^{2}$ :
\be
\sigma^x | 1 \rangle = | 0 \rangle, \quad \sigma^y | 1 \rangle = i | 0 \rangle, \quad \sigma^z | 1 \rangle = | 1 \rangle,
\lab{sigma_1} \ee
\be
\sigma^x | 0 \rangle = | 1 \rangle, \quad \sigma^y | 0 \rangle = -i | 1 \rangle, \quad \sigma^z | 0 \rangle = -| 0 \rangle.
\lab{sigma_0} \ee
The index on these symbols indicate on which $\mathbb{C}^{2}$ factor they act.

It is straightforward to see that the $z$-projection of the total spin is conserved  
\be
[H, \frac{1}{2} \: \sum_{l=0}^N (\sigma_l^z +1)]=0.
\lab{H_comm} \ee
This implies that the eigenstates of $H$ split in subspaces labeled by the number of spins over the chain that are in state $| 1 \rangle$. 

Most state transfer properties can be obtained by focusing on the 1-excitation subspace. Assume that the register is initially prepared in the state $|0\rangle ^{\otimes (N+1)} = |0,0,\dots, 0 \rangle$. Introduce the unknown state
\be
|\psi \rangle = \alpha | 0\rangle + \beta | 1\rangle \lab{psi_01} \ee
onto the first state (labeled by $l=0$). We wish to recuperate the state $|\psi \rangle$ at the last site ($l=N$) after some time. The component $\alpha |0\rangle $ is automatically found at the last site because $|0\rangle ^{\otimes (N+1)}$ is an eigenstate of $H$. 
In examining the transfer of the arbitrary state $|\psi \rangle$ it therefore suffices to consider how 
$| 1 \rangle |0\rangle ^{\otimes N}$ can possibly evolve into 
$|0\rangle ^{\otimes N} | 1 \rangle$ and that only requires looking at the one-excitation subspace. 
A natural basis for that subspace, equivalent to $\mathbb{C}^{N+1}$, is
 $$
|e_n \rangle = (0,0,\dots, 1, \dots, 0), \quad n=0,1,2,\dots,N,
$$
where the only 1 (spin up) occupies the $n$-th position. In that
basis,  the restriction $J$ of $H$ to the one-excitation subspace
is given by the following $(N+1) \times (N+1)$  Jacobi matrix
\[ J= \left( \begin{array}{ccccc}
B_{0} & J_1 & 0 & \dots  & 0 \\
J_{1} & B_{1} & J_2 & \dots  & 0 \\
\dots & \dots  & \dots & \dots &    \dots \\
 0 & 0 & \dots & J_N & B_N \end{array} \right) . \]
Its action on the basis vectors $|e_n)$ reads. \be J |e_n \rangle  =
J_{n+1} |e_{n+1} \rangle + B_n |e_n \rangle  + J_{n} |e_{n-1} \rangle .  \lab{Je} \ee Note
also that the conditions \be J_0=J_{N+1}=0 \lab{J0} \ee are
assumed.

Let $x_s,\; s=0,1,\dots, N$ be the eigenvalues of the matrix $J$.
They are all real and nondegenerate. Moreover, they are labeled in
increasing order, i.e. $x_0<x_1<x_2<\dots x_N$.

To the Jacobi matrix $J$ one can associate the monic orthogonal
polynomials  $P_n(x)$ defined by the 3-term recurrence relation
\be P_{n+1}(x) + B_n P_n(x) + U_n P_{n-1}(x) = xP_n(x), \quad
n=0,1, \dots, N,  \lab{recP} \ee where $U_n = J_n^2>0$ and
$P_{-1}=0, \; P_0(x)=1$.

$P_{N+1}(x)$ is the characteristic polynomial of the spectral
points $x_s$ \be P_{N+1}(x) = (x-x_0)(x-x_1) \dots (x-x_N).
\lab{P_N+1} \ee

The polynomials $P_n(x)$ satisfy the orthogonality relation
\be
\sum_{s=0}^N P_n(x_s) P_m(x_s) w_s = h_n \: \delta_{nm}, \lab{ort_P} \ee
where
$$
h_n = U_1 U_2 \dots U_n.
$$
The discrete weights $w_s>0$ are uniquely determined by the
recurrence coefficients $B_n, U_n$.

The perfect state transfer (PST) condition means that there exists
a time $T$ such that \be e^{iTJ} |e_0 \rangle  = e^{i \phi} |e_N \rangle,
\lab{pqc} \ee where $\phi$ is a real number. In other words, after
some time $T$, the initial state  $|e_0 \rangle $ evolves into the state
$|e_N \rangle$ (up to inessential phase factor $e^{i \phi}$).

It is well known that the PST property is equivalent to the two
conditions \cite{Kay}:

(i) the eigenvalues $x_s$ satisfy
 \be x_{s+1} - x_s=
\frac{\pi}{T} M_s, \lab{xxM} \ee where $M_s$ are positive odd
numbers.

(ii) the matrix $J$ is mirror-symmetric $RJR =J$, where the matrix $R$ (reflection matrix) is
\[ R= \left( \begin{array}{ccccc}
0 & 0 & \dots & 0  & 1 \\
0 & 0 & \dots & 1  & 0 \\
\dots & \dots  & \dots & \dots &    \dots \\
 1 & 0 & \dots & 0 & 0 \end{array} \right)\]

Property (ii) is equivalent to either two of the following properties \cite{VZ_PST}: 

(ii') the weights $w_s$ (up to a normalization) are given by the
expression \be w_s = \frac{1}{|P_{N+1}'(x_s)|}>0 . \lab{w_s_kap}
\ee 

(ii") the polynomial $P_N(x)$ satisfies 
\be
P_N(x_s) = A (-1)^s \lab{P_N_A} \ee 
with some constant $A$ not depending on $s$.

Properties (ii') and (ii") prove convenient from a constructive point of view. Indeed, a simple algorithm allows to reconstruct polynomials $P_n(x)$ and Jacobi matrix $J$ simply from a given spectrum $x_s$ verifying condition (i). As a matter of fact, the polynomials $P_{N+1}(x)$ and $P_N(x)$ can be determined explicitly using property (ii'); the polynomials $P_n(x), \: n=N-1, N-2, \dots, 1$ and the matrix $J$ are then obtained iteratively with the help of the Euclidean division algorithm.

Note that the construction ensures in all cases that the polynomials are positive definite and that the recurrence coefficients always generate a matrix $J$ with mirror symmetry.

From a mathematical perspective, interesting new families of explicit orthogonal polynomials can be generated from spectral data sets only, through this connection with chains admitting PST. A key step is in the choice of the grids. The simplest case is that of a uniform grid (i.e. $x_s$ is an arithmetic progression in $s$); in this instance formula \re{w_s_kap} gives the binomial distribution to which the Krawtchouk polynomials correspond. We thus recover the well known example described in \cite{Albanese}.

Other choices of grids $x_s$ with simple invariance properties (i.e. geometric progression, two uniform grids separated by a gap etc) were considered in \cite{VZ_PST}, \cite{VZ_HPST}.

They generate examples of orthogonal polynomials $P_n(x)$ and of corresponding Jacobi matrices $J$ with the PST property. Such explicit examples may also be useful from a practical point of view. Indeed, as was shown in \cite{VZ_PST}, starting from any explicit Hamiltonian $H$ with the PST property, one can generate many other Hamiltonians with the same PST property.     This procedure called "spectral surgery" in \cite{VZ_PST}, involves the removal of spectral points and is equivalent to Darboux transformations of the Hamiltonian $H$. Such manipulations can generate spin chains with more suitable properties (say, with a smoother behavior of the coefficients of the matrix $J$) than the initial spin chain.

\section{Finite bi-lattice and para-Krawtchouk polynomials}
\setcounter{equation}{0} Consider a finite bi-lattice of
eigenvalues \be x_s = s + \frac{1}{2}(\gamma-1)(1-(-1)^s), \quad
s=0,1,\dots,N, \lab{bi-lattice} \ee where $N$ is odd and the
parameter $\gamma$ satisfy the restrictions
$$
0 < \gamma < 2.
$$
For $\gamma \ne 1$ we have two uniform sublattices that are
intermeshed. One sublattice corresponds to even $s$: \be x_{2s} =
2s. \lab{even_x} \ee The other sublattice corresponds to odd $s$:
\be x_{2s+1} = 2s + \gamma .\lab{odd_x} \ee When $\gamma=1$ we
obtain the uniform lattice $x_s=s$. The PST condition for uniform
lattice leads to the Krawtchouk polynomials \cite{Albanese}. The
orthogonal polynomials $P_n(x)$ corresponding to the weights
\re{w_s_kap} with $x_s$ given by \re{bi-lattice} will be called
para-Krawtchouk polynomials. This term is justified by the
observation that the spectrum \re{bi-lattice} coincides (when $N =
\infty$) with the spectrum of the para-bosonic oscillator
\cite{Rosen}.

Direct calculation with the help of formula \re{w_s_kap} yields
two different expressions for the weights $w_s$ depending on
whether $s$ is even or odd: \be w_{2s} =
\frac{2^{-N}(1-\gamma/2)_J}{(1/2)_J} \: \frac{(-J)_s
(-\gamma/2-J)_s}{s! (1-\gamma/2)_s}, \quad s=0,1,\dots, J
\lab{w_even} \ee and \be w_{2s+1} =
\frac{2^{-N}(1+\gamma/2)_J}{(1/2)_J} \: \frac{(-J)_s
(\gamma/2-J)_s}{s! (1+\gamma/2)_s}, \quad s=0,1,\dots, J,
\lab{w_odd} \ee where $J=(N-1)/2$ is a positive integer (recall
that $N$ is odd).

The weights are normalized in a standard way \be \sum_{s=0}^N w_s
=1 . \lab{norm_w} \ee Equation \re{norm_w} can be verified by
observing that \be \sum_{s=0}^J w_{2s} =
\frac{2^{-N}(1+\gamma/2)_J}{(1/2)_J} \: {_2}F_1 \left(  {-J,
-\gamma/2 -J \atop 1-\gamma/2 }; 1 \right). \lab{sum_w_even} \ee
The hypergeometric function in \re{sum_w_even} can be calculated
by using the Chu-Vandermonde identity \be {_2}F_1 \left(  {-J, b
\atop c }; 1 \right) =\frac{(c-b)_J}{(c)_J} \lab{Gauss} \ee which
is valid for arbitrary $b,c$ and for positive integers $J$. Hence
\be \sum_{s=0}^J w_{2s}= \frac{2^{-N} (J+1)_J}{(1/2)_J} =1/2.
\lab{sum_even} \ee Similarly \be \sum_{s=0}^J w_{2s+1}=
\frac{2^{-N} (J+1)_J}{(1/2)_J} =1/2 \lab{sum_odd} \ee and we
obtain the desired property \re{norm_w}.

Note that the weights $w_s$ resemble the corresponding weights for
the Meixner polynomials on bi-lattices \cite{SV}.  But in contrast
to these examples of orthogonal polynomials on bi-lattices that
have been considered thus far \cite{SV}, the para-Krawtchouk
polynomials have explicit recurrence coefficients $J_n$ and $B_n$:
\be B_n = \frac{N-1+\gamma}{2}, \quad U_n =J_n^2=
\frac{n(N+1-n)((N+1-2n)^2-\gamma^2)}{4(N-2n)(N-2n+2)}.
\lab{para_ub} \ee There is an even  more convenient representation
for the recurrence coefficients \be U_n =A_{n-1} C_n , \quad B_n =
-A_n-C_n, \lab{ub_AC} \ee where \be A_n =
\frac{(N-n)(N-1-2n+\gamma)}{2(2n-N)}, \quad C_n =
\frac{n(N+1-2n-\gamma)}{2(2n-N)}. \lab{AC} \ee

The para-Krawtchouk polynomials satisfy the difference equation
\be E(x) P_n(x+2) + F(x) P_n(x-2) -(E(x)+F(x)) P_n(x) = 2n(n-N)
P_n (x), \lab{para_dfr} \ee where \be E(x) =
\frac{(x-N+1)(x-N+1-\gamma)}{2}, \quad F(x) =
\frac{x(x-\gamma)}{2}. \lab{para_EF} \ee Surprisingly, equation
\re{para_dfr} puts the para-Krawtchouk polynomials in the category
of classical OPs on uniform grids \cite{NSU}. Note however, that
the translation $x \to x+2$ corresponds to the shift $x_s \to
x_{s+2}$. This means that with respect to the grid \re{bi-lattice}
equation \re{para_dfr} is of 4-th order.

$F(x)$ vanishes at the two points $x=0$ and $x=\gamma$ which
corresponds to the smallest points of the two sub-lattices.
Similarly, $E(x)$ vanishes at the two points $x=N-1$ and
$x=N-1+\gamma$ which coincides with the largest points of the two
sub-lattices.

When $\gamma=1$ we obtain the classical symmetric Krawtchouk
polynomials with recurrence coefficients \cite{KLS} \be B_n =
\frac{N}{2}, \quad U_n = \frac{n(N+1-n)}{4}. \lab{K_ub} \ee
Equation \re{para_dfr} becomes in this case  the square of the
eigenvalue equation for the Krawtchouk polynomials \cite{KLS}.

Returning to the PST property, consider the affine transformation
of the spectral points  \re{bi-lattice} \be \t x_s = \alpha x_s +
\beta \lab{t_x} \ee with arbitrary real parameters $\alpha,\beta$.
The corresponding monic polynomials $\t P_n(x)$ will satisfy the
recurrence relation \be \t P_{n+1}(x) + \t B_n \t P_n(x) + \t U_n
\t P_{n-1}(x) = x \t P_n(x), \lab{t_P_rec} \ee where
$$
\t B_n = \alpha B_n + \beta, \quad \t U_n = \alpha^2 U_n.
$$
Using appropriate values for the parameters $\alpha, \beta$, we
can always achieve condition (i) for the PST iff \be \gamma =
\frac{M_1}{M_2}, \lab{gMM} \ee where $M_1,M_2$ are positive
co-prime integers and $M_1$ is odd.

\section{Relation with the complementary Bannai-Ito polynomials}
\setcounter{equation}{0} 
A breakthrough \cite{TVZ_BI} in the theory of orthogonal polynomials (OPs) theory has been realized recently with the discovery of classical OPs that are eigenfunctions of continuous or discrete Dunkl operators defined using reflections. These OPs are often referred to as -1 polynomials since they arise through $q \to -1$ limit of $q$-OPs. Remarkably, this has allowed for a complete characterization of the 4-parameter Bannai-Ito polynomials in a discrete variable. A significant feature of these classes of OPs is that their spectra $x_s$ depend  asymmetrically on the parity of the labeling  index as in the case of the para-Krawtchouk polynomials.

We shall now provide an explicit expression for these para-Krawtchouk polynomials in terms of the complementary Bannai-Ito polynomials \cite{TVZ_BI}. These are the kernel polynomials of the Bannai-Ito polynomials; i.e. they are obtained from the latter through a Christoffel tarnsform. These complementary Bannai-Ito polynomials depend on 4 parameters
$r_1,r_2,\rho_1, \rho_2$ and are defined through the 3-term
recurrence relation \be W_{n+1}(x) +(-1)^n \rho_2 W_n(x) + v_n
W_{n-1}(x) = x W_n(x), \lab{rec_W} \ee with \ba
&&v_{2n} = -\frac{n(n+\rho_1-r_1+1/2)(n+\rho_1-r_2+1/2)(n-r_1-r_2)}{(2n+1+g)(2n+g)}, \nonumber \\
\lab{v_BI} \\ &&v_{2n+1} =
-\frac{(n+g+1)(n+\rho_1+\rho_2+1)(n+\rho_2-r_1+1/2)(n+\rho_2-r_2+1/2)}{(2n+1+g)(2n+g+2)},
\nonumber \ea where we denote
$$
g=\rho_1+\rho_2-r_1-r_2.
$$
They can be expressed in terms of the Racah polynomials as follows
\cite{TVZ_BI} \be W_{2n}(x) = \kappa_n^{(1)} \: {_4}F_3 \left(
{-n, n+g+1, \rho_2+x, \rho_2-x  \atop  \rho_1+\rho_2+   1,
\rho_2-r_1+1/2, \rho_2-r_2+1/2 }   ; 1\right) \lab{U_Wil} \ee and
\be W_{2n+1}(x) = \kappa_n^{(2)} \: (x-\rho_2) {_4}F_3 \left( {-n,
n+g+2, \rho_2+1+x, \rho_2+1-x  \atop  \rho_1+\rho_2+   2,
\rho_2-r_1+3/2, \rho_2-r_2+3/2 }   ; 1\right), \lab{V_Wil} \ee
where the normalization coefficients $\kappa_n^{(1,2)}$ are needed
to ensure that polynomials $W_n(x)$ are monic.

Put \be r_2=\rho_2=0, \quad r_1= \frac{N+1+\gamma}{4}, \;
\rho_1=\frac{\gamma-N-3}{4}, \lab{CBI_par} \ee the recurrence
relation \re{rec_W} then reads \be W_{n+1}(x) +
\frac{n(N+1-n)((N+1-2n)^2 -\gamma^2)}{16(N-2n)(N+2-2n)} W_{n-1}(x)
= x W_n(x). \lab{rec_W_K} \ee Comparing \re{rec_W_K} with the
recurrence relation for the para-Krawtchouk polynomials  we
conclude that \be P_n(x) = 2^n \: W_n(x/2 - (N-1+\gamma)/4)
\lab{PW} \ee i.e. that the para-Krawtchouk polynomials can be
expressed in terms of the complementary Bannai-Ito polynomials
with a shifted argument.

Consider the polynomial $P_{N+1}(x)$. By assumption, $N=2M+1$ is
odd, hence by \re{U_Wil} we have \be P_{N+1}(x) =   const \;
{_3}F_2 \left( {-M-1,  x/2-M/2-\gamma/4, -x/2+M/2+\gamma/4
  \atop  \frac{\gamma-2M}{4}, -\frac{\gamma+2M}{4}}   ; 1\right) .  \lab{K_N+1} \ee
On the one hand, the hypergeometric function ${_3}F_2(1)$ in
\re{K_N+1} can be simplified using  the Pfaff-Saalsch\"utz
summation formula \cite{KLS}. This yields the factorization \be
P_{N+1}(x) = const \; \left( \frac{\gamma-x}{2}\right)_{M+1} \:
\left( \frac{x-2M}{2}\right)_{M+1}, \lab{factor_N+1} \ee where
$(x)_n =x(x=1) \dots(x+n-1)$ is the shifted factorial (Pochhammer
symbol).

On the other hand, $P_{N+1}(x)$ is the characteristic polynomial
\re{P_N+1}. Comparing \re{factor_N+1} and \re{P_N+1} we arrive at
the explicit expression \re{bi-lattice} for the spectral points.

\section{Algebraic interpretation. Hahn algebra}
\setcounter{equation}{0} It is possible to relate  the
para-Krawtchouk polynomials to the quadratic Hahn algebra
\cite{mutual}, \cite{Z_Higgs}, \cite{LV}.

Introduce two operators $X$ and $Y$ on the space of polynomials
$f(x)$: let $X$ be the operator of multiplication by the argument
$x$ : \be X f(x)=xf(x), \lab{X_def} \ee and Y be the difference
operator: \be Yf(x) = E(x) f(x+2) + F(x) f(x-2) -(E(x)+F(x)) f(x),
\lab{Y_def} \ee where the functions $E(x),F(x)$ are given by
\re{para_EF}. Note that the operator $Y$ is the operator on lhs of
equation \re{para_dfr} satisfied by the para-Krawtchouk
polynomials.

Introduce also a third operator $Z$ which is the commutator of $X$
and $Y$ \be Z = [X,Y]=XY-YX .\lab{Z_def} \ee It is then easily
verified that the  commutators $[Y,Z]$ and $[Z,X]$ are quadratic
expressions  in terms of the operators $X,Y$: \be [Y,Z]=-4\{X,Y\}
+ C_1 X + G Y + M_1, \quad  [Z,X]=-4X^2 +C_2Y +G X + M_2
\lab{rep_com} \ee where $\{X,Y\}$ stands for the anticommutator
and \ba &&C_1=4(1-N^2), \;
C_2=-4, \; G=4(N+\gamma-1), \lab{const_H} \\
&&M_1=2(N+\gamma-1)(N^2-1), \: M_2=2(1-N)(N+\gamma-1) . \nonumber
\ea The polynomial algebra with 3 generators satisfying
\re{Z_def}, \re{rep_com} as defining relations is known as the
Hahn algebra \cite{mutual}.

It has for Casimir operator \be Q =Z^2 -4 \{X^2,Y\} +(C_1 + 16)X^2
+ C_2 Y^2 + G\{X,Y\} +(2M_1-4G)X + 2M_2 Y \lab{Q_def} \ee which
commutes with the operators $X,Y,Z$: $[Q,X]=[Q,Y]=[Q,Z]=0$.

In the realization \re{X_def}, \re{Y_def} $Q$ reduces to a
constant \be Q f(x) = q f(x), \lab{Q_q} \ee where \be
q=(N-1)(N+\gamma-1)(N^2-2N+\gamma(N+3)-7). \lab{q_value} \ee  The
Hahn algebra admits a basis $\pi_n, \: n=0,1,\dots, N$, where the
operator $Y$ is diagonal \be Y \pi_n = \lambda_n \pi_n \lab{Y_pi}
\ee whereas the operator $X$ is 3-diagonal \lab{mutual} \be X
\pi_n = a_{n+1} \pi_{n+1} + B_n \pi_n + a_n \pi_{n-1}.
\lab{3diag_X} \ee The coefficients $a_n, B_n$ and $\lambda_n$ can
be calculated from the representations of the  Hahn algebra
\cite{mutual} with the initial conditions $a_0=a_{N+1}=0$: \be
\lambda_n = 2n(n-N). \lab{lambda_n} \ee The coefficients $a_n^2$
and $B_n$ coincide with the recurrence coefficients \re{para_ub}
for the para-Krawtchouk polynomials.

This gives a simple algebraic interpretation of the
para-Krawtchouk polynomials.  It is obvious that the basis vectors
$\pi_n$ are given by the para-Krawtchouk polynomials in the
realization \re{X_def}, \re{Y_def}: \be \pi_n = P_n(x). \lab{pi_P}
\ee

It is worth noting that the para-Krawtchouk polynomials can be
obtained from a periodic reduction (with period 4)  of the Darboux
transformations of finite Jacobi matrices. Indeed, as was already
noticed in \cite{SVZ_Hahn}, such periodic reductions will lead to
the Hahn algebra and to the appearance of nontrivial sub-lattices
for the spectral points.  Only ordinary Hahn polynomials were
analyzed from this point of view in \cite{SVZ_Hahn}. The
para-Krawtchouk polynomials provide a new non-trivial example of
such periodic closures.

\section{The case of odd number of the eigenvalues}
\setcounter{equation}{0} So far, we considered the case of an even
number $N+1$ of eigenvalues $x_s, \: s=0,1,\dots, N$  with $N=1,3,
5\dots$. Using the Christoffel transform, it is possible to obtain
the para-Krawtchouk polynomials corresponding to an odd number of
eigenvalues $x_s, \: s=0,1,\dots,N-1$. Indeed, consider the
polynomials \be \t P_n(x) = \frac{P_{n+1}(x) - A_n
P_n(x)}{x-x_{N}}, \lab{CTP} \ee where $A_n$ are given by \re{AC}.
It is not difficult to check that
$$
A_n = \frac{P_{n+1}(x_N)}{P_n(x_N)}
$$
and hence formula \re{CTP} defines new orthogonal polynomials $\t
P_n(x)$ obtained by the  Christoffel transform \cite{Sz} of the
polynomials $P_n(x)$. The polynomials $\t P_n(x)$ satisfy the
recurrence relation \be \t P_{n+1}(x) + \t B_n \t P_n(x) + \t U_n
\t P_{n-1}(x) = x \t P_n(x) \lab{rec_CP} \ee with the coefficients
\cite{VZ_PST} \be \t U_n = U_n \frac{A_n}{A_{n-1}}, \quad \t B_n =
B_{n+1} +A_{n+1} - A_n . \lab{CT_ub} \ee A simple calculation
yields \be \t U_n = \frac{n(N-n)((2n-N)^2-(\gamma-1)^2)}{(2n-N)^2}
\lab{tu} \ee and \be \t B_n = \frac{N+\gamma}{2}-1 +
\frac{(\gamma-1)N}{4} \left(\frac{1}{2n-N} - \frac{1}{2n+2-N}
\right) . \lab{tb} \ee The Jacobi matrix $\t J$ with the
recurrence coefficients $\t U_n, \t B_n$ has size $N \times N$ and
its eigenvalues $\t x_s$ coincide with the eigenvalues of the
matrix $J$ apart from the final value $x_N$, i.e. \be \t x_s = s +
\frac{1}{2}(\gamma-1)(1-(-1)^s), \quad s=0,1,\dots,N-1 . \lab{txs}
\ee Thus the polynomials $\t P_n(x)$ defined by \re{rec_CP},
\re{tu}, \re{tb} have properties similar to those of the
para-Krawtchouk polynomials. Clearly, the Jacobi matrix $\t J$
generates a spin chain with PST property under the  restriction
\re{gMM}. Indeed, Christoffel transforms like \re{CTP} preserve
the PST property and can be used to construct new examples of spin
chains with PST (see \cite{VZ_PST} for details).

\section{Conclusions}
\setcounter{equation}{0}
In summary, we have presented an analytically solvable XX spin chain model with
inhomogeneous nearest-neighbor couplings that effect perfect state transfer. It was
obtained by exploiting the intimate connection between inverse spectral problems for
Jacobi matrices and orthogonal polynomial theory. We posited a spectrum of bi-lattice form for 1-excitations  and identified the corresponding coupling constants of the
Hamiltonian. The associated orthogonal polynomials were seen to form a remarkable
family that had hitherto not been identified. They have been called para-Krawtchouk
polynomials as their spectral properties are analogous to that of the parabose oscillator.
We have presented their characterization and observed that they enjoy properties similar
to those of classical OPs. We have further shown that they are related to the quadratic
Hahn algebra.

We have thus provided an example of fruitful cross-fertilization between the study
of systems with PST and the theory of orthogonal polynomials. We trust further results
can be reaped from this connection and in view of their nice features, we believe that
the para-Krawtchouk polynomials will find more applications now that they have been
discovered and characterized.

\newpage

\bigskip\bigskip
{\Large\bf Acknowledgments}
\bigskip

AZ thanks Centre de Recherches Math\'ematiques (Universit\'e de
Montr\'eal) for hospitality.  The authors would like to thank M.
Christandl and V. Spiridonov for stimulating discussions. The
research of LV is supported in part by a research grant from the
Natural Sciences and Engineering Research Council (NSERC) of
Canada.
\newpage

\bb{99}

\bi{Albanese} C.Albanese, M.Christandl, N.Datta, A.Ekert,  {\it
Mirror inversion of quantum states in linear registers}, Phys.
Rev. Lett. {\bf 93} (2004), 230502;  arXiv:quant-ph/0405029

%\bi{BG} C. de Boor and G.~H.~Golub {\it The numerically stable reconstruction of a Jacobi matrix from spectral data}, Lin. %Alg. Appl. {\bf 21} (1978), 245--260.

%\bi{BS} C. de Boor, E.Saff, {\it Finite sequences of orthogonal polynomials connected by a Jacobi matrix}, Lin. Alg. Appl. %{\bf 75} (1986), 43--55

%\bi{Bor} A.Borodin, {\it Duality of Orthogonal Polynomials on a Finite Set}, J. Stat. Phys.
%{\bf 109}, (2002), 1109--1120.

%\bi{Chi} T. Chihara, {\it An Introduction to Orthogonal
%Polynomials}, Gordon and Breach, NY, 1978.

\bi{mutual} Ya.I.Granovskii, I.M.Lutzenko and A.S.Zhedanov,  {\it
Mutual integrability, quadratic algebras and dynamical
symmetry}, Ann.Phys. (USA) {\bf 217} (1992), 1--20.

%\bi{Ismail_book} M.E.H.Ismail, {\it Classical and Quantum orthogonal polynomials in one variable}.
%Encyclopedia of Mathematics and its Applications (No. 98), Cambridge, 2005.

%\bi{JSJ} Jafarov, Stoilova, Van der Jeugt, {Finite oscillator models: the Hahn oscillator}, ArXiv: 1101.5310.

%\bi{JSJ2} Jafarov, {\it The ${\mathfrak{su}}(2)_{\alpha}$ Hahn
%oscillator and a discrete Hahn-Fourier transform}

%\bi{JJ} H.~Jeffreys and B.~S.~Jeffreys, {\it Methods of Mathematical Physics}, 3rd ed. Cambridge, England: Cambridge %University Press, 1988.

\bi{Kay} A.Kay, {\it A Review of Perfect State Transfer and its
Application as a Constructive Tool}, Int. J. Quantum Inf. {\bf 8}
(2010), 641--676;  arXiv:0903.4274.

\bi{KLS} R. Koekoek,P. Lesky, R. Swarttouw, {\it Hypergeometric Orthogonal Polynomials and Their Q-analogues}, Springer-Verlag, 2010.

\bi{LV} P. L\'etourneau and  L. Vinet, {\it Superintegrable
Systems: Polynomial Algebras and Quasi-Exactly Solvable
Hamiltonians}, Ann. Phys. {\bf 243} (1995), 144--168.

\bi{NSU} A.F. Nikiforov, S.K. Suslov, and V.B. Uvarov, {\em
Classical Orthogonal Polynomials of a Discrete Variable},
Springer, Berlin, 1991.

%\bi{Shi} T. Shi, Y.Li , A.Song, C.P.Sun, {\it Quantum-state transfer via the ferromagnetic chain in a spatially modulated
%field}, Phys. Rev. {\bf A 71} (2005), 032309, 5 pages, quant-ph/0408152.

\bi{SV} C.Smet and W.Van Assche, {\it Orthogonal polynomials on a bi-lattice}, Construct. Approx.
{\bf 36} (2012), 215--242;  arXiv:1101.1817 math.

%\bi{SJ} N.Stoilova, J.Van der Jeugt, {\it An exactly solvable spin chain related to Hahn polynomials}, SIGMA {\bf 7} (2011), %033, 13 pages.

\bi{Rosen} M. Rosenblum, {\it Generalized Hermite Polynomials and
the Bose-like Oscillator Calculus}, in: Oper. Theory Adv. Appl.,
vol. {\bf 73}, Birkhauser, Basel, 1994, pp. 369--396.
ArXiv:math/9307224.

\bi{SVZ_Hahn} V.Spiridonov, L.Vinet, A.Zhedanov, {\it Periodic
reduction of the factorization chain and the Hahn polynomials}, J.
Phys. A: Math. Gen. {\bf 27} (1994), L669--L675.

\bibitem{Sz} G. Szeg\H{o}, Orthogonal Polynomials, fourth edition,  AMS, 1975.

\bi{TVZ_BI} S.Tsujimoto, L.Vinet and A.Zhedanov, {\it Dunkl shift operators and Bannai-Ito polynomials}, Adv.
Math. {\bf 229} (2012) 2123-–58 arXiv:1106.3512.

%\bi{TVZ_para} S.Tsujimoto, L.Vinet and A.Zhedanov, {\it From
%$sl_q(2)$ to a Parabosonic Hopf Algebra},  SIGMA {\bf 7} (2011),
%093, 13 pages; arXiv:1108.1603.

%\bi{TVZ_Hahn} S.Tsujimoto, L.Vinet and A.Zhedanov, {\it Dual -1
%Hahn polynomials:  "classical" polynomials beyond the Leonard
%duality}, arXiv:1108.0132 (to be published in Proc.Amer.Math.Soc.)

%\bi{Wang} Y.~Wang, F.~Shuang, H.~Rabitz, {\it All possible coupling schemes in XY spin chains for perfect state transfer}, %Phys. Rev. {\bf A 84}, (2011) 012307, arXiv:1101.1156.

\bi{VZ_PST} L.Vinet and A.Zhedanov, {\it How to construct spin
chains with perfect state transfer}, Phys. Rev. {\bf A 85} (2012), 012323 arXiv:1110.6474.

\bi{VZ_HPST} L.Vinet and A.Zhedanov, {\it Dual -1 Hahn polynomials and perfect state transfer}, arXiv:1110.6477,

\bi{Z_Higgs} A.Zhedanov, {\it The "Higgs algebra" as a quantum deformation of $su(2)$}, Mod.Phys.Let. {\bf A7} (1992), 507--512.

\eb

\end{document}